\documentclass[showpacs,twocolumn,prb]{revtex4-1}

\usepackage{graphicx, amsmath, amsthm, amssymb,graphics}

\newcommand{\be}{\begin{equation}}
\newcommand{\ee}{\end{equation}}
\newcommand{\bea}{\begin{eqnarray}}
\newcommand{\eea}{\end{eqnarray}}
\newcommand{\bwt}{\begin{widetext}}
\newcommand{\ewt}{\end{widetext}}

\newcommand{\bff}{\rm}

\newcommand{\up}{\uparrow}
\newcommand{\dn}{\downarrow}

\begin{document}

\title{Optical conductivity for a dimer in the Dynamic Hubbard model}
\author{G. H. Bach and F. Marsiglio}
\affiliation{Department of Physics, University of Alberta, Edmonton, Alberta,
Canada, T6G~2G7}

\begin{abstract}
The Dynamic Hubbard Model represents the physics of a multi-band Hubbard model by using a pseudo-spin degree of freedom to dynamically modify the on-site Coulomb interaction. Here we use a dimer system to obtain analytical results for this model. The spectral function and the optical conductivity are calculated analytically for any number of electrons, and the distribution of optical spectral weight is analyzed in great detail. The impact of polaron-like effects due to overlaps between pseudo-spin states on the optical spectral weight distribution is derived analytically. Our conclusions support results obtained previously with different  models and techniques: holes are less mobile than electrons.
\end{abstract}

\date{\today}
\maketitle

\section{Introduction}
The occurrence of electron-hole asymmetry in tunneling spectra,\cite{pasupathy08} and the anomalous behaviour
in the optical conductivity sum rule at the superconducting transition 
temperature\cite{basov99,santander01,molegraaf02,carbone06,vanheumen07} both contribute to the possibility that the superconductivity in the cuprate materials is unusual in several respects. In particular, the notion of `kinetic energy-driven superconductivity' is implied by the optical experiments, as predicted almost ten years in advance of these experiments.\cite{hirsch92}

{\bff These experiments indicated that a significant transfer of spectral weight occurs in the cuprates,\cite{charnukha11} both in the normal state as a function of temperature, and as
a result of the superconducting transition. More importantly, perhaps, is the range of frequencies affected by the transition, as a significant amount of spectral weight is transferred from
very high frequencies to very low frequencies. Thus, there is an apparent violation of the Ferrell-Glover-Tinkham optical sum rule, as an examination of the low frequency region alone shows a spectral weight discrepancy. This indicates that physics beyond the usual paradigm of ``energy lowering due to potential energy considerations'' is at work; in particularly the anomalous sign of the change in low frequency optical spectral weight indicates that some mechanism involving the kinetic energy of the carriers is at work.}

{\bff Earlier modelling of cuprate superconductivity\cite{hirsch89c} includes some of this physics --- this is what motivated the relatively early theoretical discussion of optical spectral weight transfer} --- but a more recent theoretical model, advanced more than ten years ago\cite{hirsch01} goes further to explain some of the anomalous features in the spectroscopic measurements of the cuprates; this is the so called Dynamic Hubbard model. This model utilizes a phenomenological pseudo-spin degree of freedom at each lattice site designed to mimic orbital relaxation effects which necessarily occur in real atoms. {\bff As far as optical spectral weight transfer is concerned, this model includes higher frequency excitations (here modelled by the pseudo-spins), and therefore, while we do not address superconductivity or temperature effects in this paper, by using the Dynamical Hubbard model, we can study how spectral weight transfer occurs as a function of doping.}

The Hamiltonian for the Dynamical Hubbard model is\cite{hirsch01}
\bea
H_{\rm DHB}= -t \ \sum_{\langle i,j \rangle \sigma}(c^{\dagger}_{i\sigma}c_{j\sigma}+c^{\dagger}_{j\sigma}c_{i\sigma}) - \mu \sum_{i,\sigma}n_{i\sigma} \nonumber\\
+ \sum_{i}(\omega_{0}\sigma^{x}_{i}+ g\omega_{0}\sigma^{z}_{i})
+ \sum_{i}(U-2g\omega_{0}\sigma^{z}_{i})n_{i\up}n_{i\dn}
\label{ham_DHB}
\eea
where the pseudo-spin degree of freedom is here represented by a Pauli operator $\sigma_i$ at each site; it interacts with the electron charge through the double occupancy term, and contributes a dynamical interaction in addition to the usual Hubbard interaction. The rest of the Hamiltonian is as follows:
the first term represents the electron kinetic energy within a tight-binding model with one orbital per site. {\bff Note that we are really trying to understand physics that originates in processes involving multiple orbitals. It is desirable to minimize the complexity by retaining a single orbital, and it is then the pseudo-spin that acts to mimic the physics of carriers undergoing transitions between multiple orbitals when the local occupation changes.}The second term determines the electron density through the chemical potential, the third term defines the two level system for the 
pseudo-spin degree of freedom at each site, and the last term is the on-site interaction which, in addition to the short range Coulomb repulsion represented by $U$, is also modulated through a coupling constant $g \omega_0$ by the state of the pseudo-spin. 

When the double occupancy is high, the pseudo-spin will reside in its excited state for the sake of minimizing the Coulomb repulsion, much like the phenomenon in real atoms, where two electrons will sacrifice having a minimal electron-ion energy and spread out amongst the excited orbitals in order to minimize their Coulomb repulsion. In the opposite limit, when the double occupancy is very low, electrons will tend to stay in the lowest energy state available in the given atom (loosely, the Wannier state which is being modelled in the tight-binding Hamiltonian), and the Coulomb energy will be high, though irrelevant, since only rarely will two electrons occupy the same site.

The Dynamic Hubbard model contains at least some of the phenomenology of hole 
superconductivity,\cite{hirsch89a,hirsch89b,hirsch89c,marsiglio90} proposed more than twenty years ago. In particular, the model contains electron-hole asymmetry, where holes at the top of a band are heavier than electrons at the bottom of a band. Electron-hole asymmetry can arise in a number of ways: just having further than nearest neighbour hopping can result in a band mass asymmetry, upon which polaronic mass asymmetry can build.\cite{chakraborty11} Even with just nearest neighbour hopping, lattice geometry can also result in electron-hole asymmetry, and again be amplified through polaron effects.\cite{hague06} Here, the electron-hole asymmetry arises through realistic interactions, and is strongly connected to the fact that holes pair more readily than electrons by lowering their kinetic energy in the superconducting phase, a phenomenon supported by the anomalous observations in the optical conductivity sum rule mentioned above. More recently, the Dynamic Hubbard model has been explored with Dynamical Mean Field Theory (DMFT).\cite{bach10} In particular, the spectral function and the optical conductivity were calculated, in the normal state, to illustrate the electron-hole asymmetry present in the model.

In this paper we will focus on the optical conductivity, and provide a complementary calculation involving a simple dimer. Such a small system does not constitute a very realistic system; however, dimer calculations have an illustrious history for providing insight into models.\cite{harris67,ranninger92,avella03} Furthermore, following Ref. [\onlinecite{hirsch92}], a dimer calculation provides good insight into the processes that contribute to the conductivity. We will briefly review the optical conductivity and the sum rule  in the next section, and describe the details of the dimer calculation in the third section, in perturbation theory; we have also performed exact diagonalizations to delineate the regime of validity of the perturbative results. This is followed with a discussion of the results, particularly in light of the DMFT results reported in Ref.~[\onlinecite{bach10}]. We then conclude with a summary.

\section{Optical conductivity and related sum rules}

The real part of the optical conductivity $\sigma_1(\omega)$, as a function of frequency, $\omega$,  at finite temperature $T$ can be written as:
\bea
\sigma_1(\omega)=\frac{\pi}{Z} \sum_{n,m} \frac{e^{-\beta E_n}-e^{-\beta E_m}}{E_m-E_n}|\langle m|{J}|n\rangle|^2\nonumber \\
\delta(\omega-\frac{E_m-E_n}{\hbar})
\eea
where $|m\rangle$ and $|n\rangle$ are eigenstates of the Hamiltonian of the system, $Z$ is the partition function, 
$\beta=1/k_BT$ and ${J}$ is the current operator, obtained through the polarization operator.\cite{mahan00} As formulated by Kubo in 1957,\cite{kubo57} the optical conductivity satisfies the general sum rule,  
\bea
\int^{+\infty}_{0} \rm{Re}\,[\sigma_{\mu\nu}(\omega)]d\omega=\frac{\pi}{2}\sum_r \frac{e^2_r n_r}{m_r} \delta_{\mu\nu}
\label{srule1}
\eea
where $r$ denotes the type of charge carrier, and $n_r$, $e_r$ and $m_r$ are the number density, charge, and mass, respectively, of the $r$-type carrier, and $\mu,\nu$ are the indices of the conductivity tensor. For an isotropic electron system, the sum rule (\ref{srule1}) is rewritten as:
\bea
\int^{+\infty}_{0} \sigma_1(\omega)d\omega=\frac{\pi e^2n}{2m},
\label{srule2}
\eea
with $m$ the bare mass of electrons and $n$ the total electron density.\cite{smith78}

In condensed matter systems we often work with effective Hamiltonians, for example formulated for a single band within tight-binding. One can then formulate a sum rule restricted to that single band, and obtain\cite{maldague77,millis04,marsiglio08}
\bea
\int^{+\infty}_{0} Re[\sigma_{xx}(\omega)]d\omega=\frac{\pi e^2}{4\hbar^2}\Bigg\{\frac{4}{N}\sum_k \frac{\partial^2\epsilon_k}{\partial k^2_x}n_k \Bigg\},
\label{srule3}
\eea
where $n_k$ is the single electron occupation number, and $\epsilon_k$ is the dispersion relation for the non-interacting electrons. In reality, the integration in Eq.~(\ref{srule3}) is taken up to a cut-off frequency $\omega_c$ determined experimentally in order to allow only intraband transitions, and to avoid the inevitable interband transitions which are {\em not} part of the sum rule Eq.~(\ref{srule3}). Theoretically, the right-hand-side (RHS) of Eq.~(\ref{srule3}) is often used,\cite{norman07} as this is much simpler to calculate. Furthermore, when only nearest neighbour hopping is allowed on a hypercubic lattice, the sum rule Eq.~(\ref{srule3}) reduces to:\cite{maldague77}
\bea
\int^{\omega_c}_{0} \sigma_1(\omega) d\omega=-\frac{\pi e^2a^2}{2\hbar^2}\langle{K}\rangle,
\label{srule4}
\eea
with ${K}=-t\sum_{\langle ij \rangle,\sigma=\up\dn}(c^{\dagger}_{i\sigma}c_{j\sigma}+h.c.)$ and $a$ the lattice constant. From the RHS of Eq.~(\ref{srule4}), it is clear that the optical sum depends not only on the external parameters (such as the temperature) but also on the electronic structure of the system. The validity of using the kinetic energy on the RHS instead of the expression in Eq.~(\ref{srule3}), even when the dispersion is not just nearest neighbour hopping {\bff has been explored in Refs.~[\onlinecite{marsiglio06} and \onlinecite{toschi08}],} to which the reader is referred. In the following we assume Eq.~(\ref{srule4}) holds.

Measurements of the optical conductivity sum rule in a number of the cuprate superconductors generally show an increase of spectral weight in the low frequency regime in the superconducting state, at least in the underdoped and optimally doped materials.\cite{santander01,molegraaf02,carbone06,vanheumen07} This enhancement of the optical sum at the superconducting transition temperature conflicts with the result from BCS-like superconductivity where an increase in the kinetic energy (and therefore a decrease in the single band optical sum) is expected instead. 
This can be explained through a number of different scenarios, examples of which are preformed pairs\cite{alexandrov94}, and phase fluctuations.\cite{eckl03,toschi05,kyung06} {\bff In contrast to the model considered here, many of these calculations have, as a key ingredient, proximity to a nearby Mott insulating state.\cite{toschi05,haule07} Note that some authors\cite{karakozov02,norman07} have attributed the anomalous temperature dependence of the low frequency optical spectral weight to a cutoff effect (required in the experimental analysis). Karakozov et al.\cite{karakozov02} have also attributed the anomalous change in spectral weight at the superconducting transition temperature to a cutoff effect, though this has been refuted in Ref.~[\onlinecite{marsiglio08}]. See Ref.~[\onlinecite{marsiglio09}] for a brief review.}

On the other hand, models like the hole mechanism of ``kinetic energy driven" superconductivity support the idea of minimizing the total energy by reducing the kinetic energy, and therefore the optical sum has an anomalous temperature dependence below $T_c$.\cite{hirsch02_science} When the system goes superconducting, the missing optical spectral weight  is predicted to be distributed over the whole range of frequencies, i.e. weight is transferred from high frequency to low frequency and the low energy sum rule appears to increase as a consequence.\cite{hirsch92,hirsch00} 
\section{Optical conductivity in a dimer}

We proceed now with a brief discussion of the site Hamiltonian followed by a detailed description of the dimer.

\subsection{Properties of the Hamiltonian}

Following Refs.~[\onlinecite{hirsch02b,bach10}], we begin with the on-site Hamiltonian {\em for electrons}: 
\bea
H^{(i)}_{\rm{DHM}} = \omega_{0}\sigma^{i}_{x} + g\omega_{0}\sigma^{i}_{z} +[U-2g\omega_{0}\sigma^{i}_{z}]n_{i\up}n_{i\dn}.
\label{ham_onsite}
\eea
The solutions are provided in detail in Refs. [\onlinecite{hirsch02b,bach10}]; for $n$ electrons the ground state ($|n\rangle$) and the first excited state ($|\bar{n}\rangle$) are
\bea
|n\rangle&=&u(n)|+\rangle-v(n)|-\rangle \nonumber \\
|\bar{n}\rangle&=&v(n)|+\rangle+u(n)|-\rangle
\label{eigenstates}
\eea
with 
\bea
u(0)&=&u(1)=v(2) \nonumber \\
v(0)&=&v(1)=u(2),
\label{comp}
\eea
and
\bea
u(0)=\sqrt{\frac{1}{2}\biggl(1-\frac{g}{\sqrt{1+g^2}}\biggl)} \nonumber \\
v(0)=\sqrt{\frac{1}{2}\biggl(1+\frac{g}{\sqrt{1+g^2}}\biggl)}.
\label{uandv}
\eea
The eigenvalues (ground state $\epsilon(n)$ and excited state $\bar{\epsilon}(n)$) are
\bea
\epsilon(n) &=& \delta_{n,2}U -\omega_{0}\sqrt{1+g^2}
\nonumber \\
\bar{\epsilon}(n) &=& \delta_{n,2}U +\omega_{0}\sqrt{1+g^2}.
\label{eigenvalues}
\eea

Especially important for the hopping processes is the overlap of background spin states with different numbers of electron; these are 
\bea
T &=&  \langle0|1\rangle= u(0)u(1)+v(0)v(1) \nonumber \\
    &=& u(0)^2 + v(0)^2 =1 \nonumber \\
S &=& \langle1|2\rangle  =  u(1)u(2)+v(1)v(2) \nonumber \\
    &=& 2u(1)v(1) =  \frac{1}{\sqrt{1+g^2}}.
\label{overlap}
\eea
These parameters play an important role for the spectral function; these are defined in Ref.~[\onlinecite{hirsch02c}] as $A_{n+1,n}$ for electron destruction in a system of $n+1$ electrons (and $A_{n,n+1}$ for electron creation in a system of $n$ electrons. For a single site these single particle spectral functions are\cite{hirsch02c}
\bea
A_{10}(\omega)&=&A_{01}(\omega)= \delta(\omega) \nonumber \\
A_{12}(\omega)&=&A_{21}(\omega)=S^2\delta(\omega)+(1-S^2)\delta(\omega-\Omega_0).
\label{spectral}
\eea
Even though the spectral weight is calculated for a single site, it is clear that there is a reduction of the weight at zero frequency if the second electron is added to the one-electron ground state. The reason is because there are two possibilities: the pseudo-spin can remain in the same state as the first electron with a probability $S^2<1$, or it can become excited with an energy cost $\Omega_0 = 2\omega_0 \sqrt{1 + g^2}$. In the thermodynamic limit, this effect is known as the reduction of quasiparticle weight by transferring part of the coherent contribution (at $\omega = 0$) to the incoherent part (at large $\omega$), resulting in a one particle spectral weight, $z<1$. Since the quasiparticle weight is inversely proportional to the effective mass, $z \sim m/m^\star$, this statement means that holes are heavier (or more `dressed'\cite{hirsch00b}) than electrons.

For calculating the optical conductivity in perturbation theory, the Hamiltonian is divided into two parts:
\bea
H &=& H_0 + H^\prime, \\
H_0 &=& \sum_{i} (\omega_{0}\sigma^{x}_{i}+ g\omega_{0}\sigma^{z}_{i}) + (U-2g\omega_{0}\sigma^{z}_{i})n_{i\up}n_{i\dn} \\
H^\prime &=& K =    -t \ \sum_{\langle i,j \rangle \sigma}(c^{\dagger}_{i\sigma}c_{j\sigma}+c^{\dagger}_{j\sigma}c_{i\sigma}),
\eea
where $H_0$ is the site Hamiltonian and $H^\prime$ is the hopping part which is considered as a perturbation under the following conditions. Based on the definition of the overlaps between the pseudo-spin ground states in Eq.~(\ref{overlap}), we can define a pseudo-spin state for a given number of electrons in terms of the eigenstates involving
a different number of electrons. These overlaps contain the background deformations (modelled by the pseudo-spin)
that must be `dragged' along as the electron hops. Thus, following Ref.~[\onlinecite{hirsch92}]:  
\bea
|1\rangle&=&S|2\rangle-\bar{S}|\bar{2}\rangle\\
|\bar{1}\rangle&=&\bar{S}|2\rangle + S|\bar{2}\rangle\\
|2\rangle&= &S|1\rangle +\bar{S}|\bar{1}\rangle\\
|\bar{2}\rangle&=&-\bar{S}|1\rangle+S|\bar{1}\rangle
\eea
where
\be
\bar{S} = \sqrt{1-S^2}=\frac{g}{\sqrt{1+g^2}}.
\ee
We wish to solve this problem in all number sectors (one, two, and three electrons). We cover these in the following subsections.

\subsection{Three electron sector}

Beginning with
three electrons on the dimer (a hole-like configuration), the (non-normalized) ground state wave function is given in first order perturbation theory as:
\bea
|\psi^{(3)}_0\rangle=|1\rangle_o + \frac{\sqrt{2}tS\bar{S}}{\Omega_0} |3\rangle_o - \frac{t\bar{S}^2}{2\Omega_0} |4\rangle_o
\eea
where
\bea
|1\rangle_{e \atop o}&=& \frac{1}{\sqrt{2}}[c^{\dagger}_{a\up}c^{\dagger}_{a\dn}|2\rangle_a\otimes c^{\dagger}_{b\up}|1\rangle_b 
\pm c^{\dagger}_{a\up}|1\rangle_a \otimes c^{\dagger}_{b\up}c^{\dagger}_{b\dn}|2\rangle_b]  \nonumber \\
|2\rangle_{e \atop o} &=&\frac{1}{2} [\pm c^{\dagger}_{a\up}|1\rangle_{a} \otimes c^{\dagger}_{b\up}c^{\dagger}_{b\dn}|\bar{2}\rangle_{b} 
\pm c^{\dagger}_{a\up}|\bar{1}\rangle_{a} \otimes c^{\dagger}_{b\up}c^{\dagger}_{b\dn}|2\rangle_{b}  \nonumber\\
&+& c^{\dagger}_{a\up}c^{\dagger}_{a\dn}|2\rangle_a \otimes c^{\dagger}_{b\up}|\bar{1}\rangle_b+c^{\dagger}_{a\up}c^{\dagger}_{a\dn}|\bar{2}\rangle_a \otimes c^{\dagger}_{b\up}|1\rangle_b] \nonumber \\
|3\rangle_{e \atop o} &=&\frac{1}{2} [\mp c^{\dagger}_{a\up}|1\rangle_{a} \otimes c^{\dagger}_{b\up}c^{\dagger}_{b\dn}|\bar{2}\rangle_{b} 
\mp c^{\dagger}_{a\up}|\bar{1}\rangle_{a} \otimes c^{\dagger}_{b\up}c^{\dagger}_{b\dn}|2\rangle_{b}  \nonumber\\
&+& c^{\dagger}_{a\up}c^{\dagger}_{a\dn}|2\rangle_a \otimes c^{\dagger}_{b\up}|\bar{1}\rangle_b+c^{\dagger}_{a\up}c^{\dagger}_{a\dn}|\bar{2}\rangle_a \otimes c^{\dagger}_{b\up}|1\rangle_b] \nonumber \\
|4\rangle_{e \atop o} &=& \frac{1}{\sqrt{2}}[c^{\dagger}_{a\up}c^{\dagger}_{a\dn}|\bar{2}\rangle_a\otimes c^{\dagger}_{b\up}|\bar{1}\rangle_b
\pm c^{\dagger}_{a\up}|\bar{1}\rangle_a \otimes c^{\dagger}_{b\up}c^{\dagger}_{b\dn}|\bar{2}\rangle_b],\nonumber\\
\label{3states}
\eea
are the 8 basis states required to span the Hilbert space for the three electron sector.
Here the subscript `$e$' refers to the even states, and $a$ and $b$ are the indices of the first and second site, respectively, in the dimer. Note that kets followed by the subscript $a$ or $b$ have numbers 0, 1, or 2 (with or without bars on top) that refer to the pseudo-spin eigenstates defined in Eq.~(\ref{eigenstates}), whereas kets followed by $e$ (for `even') or $o$ (for `odd') refer to linear combinations of product states of electrons and pseudo-spin eigenstates, as, for example, in Eq.~(\ref{3states}).

\begin{figure}[tp]
\centering
\includegraphics[height=2.7in,width=3.in]{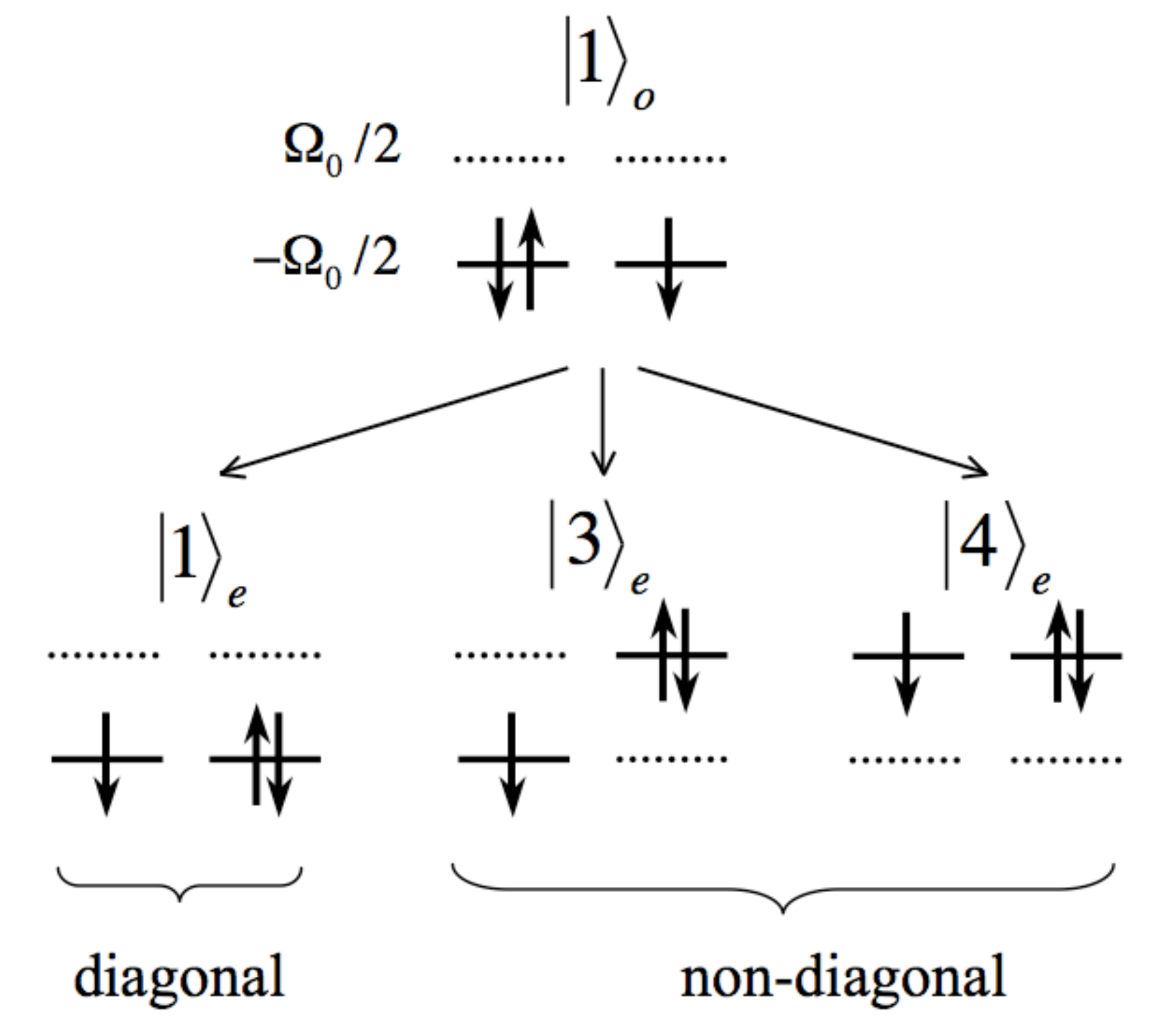}
\caption{Schematic depiction of optical transitions in a dimer with three electrons. The lines (both solid and {dashed}) show the two available levels of the pseudo-spin energy at each site; the solid lines correspond to the occupied pseudo-spin state and the {dashed} lines {correspond to} the unoccupied state. Transitions between states with the same pseudo-spin energy levels are diagonal; these contribute to the intraband conductivity, while non-diagonal transitions between states with different pseudo-spin energy levels modify the interband conductivity. State labels are those found in Eq.~(\ref{3states}) in the text, where they are given in full even or odd form.}
\end{figure}

The three-particle ground state energy, to first order in the hopping perturbation, is given by:
\be
E^{(3)}_0 \equiv {_{o}\langle} 1 | H | 1 \rangle_{o} = U-tS^2-\Omega_0,
\label{groundstate_energy}
\ee
and the excited state energies for the three electron sector, are,
\bea
E^{(3)}_1 &\equiv& {_{e}\langle} 1 | H | 1 \rangle_{e} = U +tS^2 -\Omega_0 \nonumber \\
E^{(3)}_2 &\equiv& {_{e \atop o}\langle} 2 | H_0 | 2 \rangle_{e \atop o} = U \nonumber \\
E^{(3)}_3 &\equiv& {_{e \atop o}\langle} 3 | H_0 | 3 \rangle_{e \atop o} = U \nonumber \\
E^{(3)}_4 &\equiv& {_{e \atop o}\langle} 4 | H_0 | 4 \rangle_{e \atop o} = U + \Omega_0.
\label{zeroth_energies}
\eea
Note that there are degeneracies at zeroth order between even and odd states; these are broken in first order perturbation theory, as is explicitly written in Eq.~(\ref{groundstate_energy}) and the first of Eqs. (\ref{zeroth_energies}).
Only the zeroth order energies (and wave functions) are needed for the other excited states, and that is what is written here. Also, the Hamiltonian will only couple states of a given parity, whereas the conductivity will couple only states of opposite parity.

The optical conductivity for the dimer at zero temperature can be calculated for the three electron sector as:
\bea
\sigma_1(\omega)=\pi \sum_{m \ne 0} \frac{|\langle \psi^{(3)}_0 |{J}| m \rangle_e|^2}{E_m^{(3)}-E_0^{(3)}} \delta(\omega-\frac{E_m^{(3)} - E_0^{(3)}}{\hbar})
\label{conductivity}
\eea
where $| m \rangle_e$  are the excited states of the system (only even parity is required since the ground state has odd parity --- these are given in Eq.~(\ref{3states})), and
 \bea
 {J}= \frac{iet}{\hbar} \sum_{\sigma} (c^{\dagger}_{a\sigma}c_{b\sigma}-c^{\dagger}_{b\sigma}c_{a\sigma})
 \eea
is the current operator. By acting with the ${J}$ operator on the ground state $|\psi^{(3)}_0\rangle$, we connect to three of the excited states (all even parity) given in Eq.~(\ref{3states}). Note that since the current operator is already of order $t$, only zeroth order wave functions are required, but for the first even parity excited state, first order corrections to the energy are required in the denominator of Eq. (\ref{conductivity}) to break the degeneracy.

Operating with the current operator on the unperturbed ground state gives
\be
{J} | 1\rangle_0 = {iet \over \hbar} \biggl( S^2|1\rangle_e - \sqrt{2} S\bar{S} | 3\rangle_e - \bar{S}^2|4\rangle_e  \biggr)
\label{current}
\ee
so that the optical conductivity for three electrons includes three peaks. These three transitions are shown schematically in Fig.~1; the analytical expression for the optical conductivity (for three electrons) is
\bea
\sigma_{1}^{(3)}(\omega) &=& \frac{\pi e^2 t}{2\hbar^2}[S^2\delta(\omega-\frac{2tS^2}{\hbar})+4S^2{\bar S}^2
\frac{t}{\Omega_0}\delta(\omega-\frac{\Omega_0}{\hbar}) \nonumber \\
&+& {\bar S}^4\frac{t}{\Omega_0} \delta(\omega-\frac{2\Omega_0}{\hbar})].
\label{opt1}
\eea
The optical sum rule can be checked by calculating the expectation value of the ${K}$ operator in the ground state,
\be
\langle\psi^{(3)}_{0}|-{K}|\psi^{(3)}_{0}\rangle = 
t
\biggl(S^2 + 4S^2\bar{S}^2\frac{t}{\Omega_0}+\bar{S}^4 \frac{t}{\Omega_0}\biggr),
\ee
which is precisely the combination of weights given in Eq.~(\ref{opt1}). The first contribution comes from intraband transitions --- this would correspond to the Drude weight for an extended system. This Drude response is, however, weighted by the overlap $S=\langle1|2\rangle$ between the respective ground states of the pseudo-spin with one and two electrons. This is referred to as a `diagonal' transition in Fig.~1, since the background (here, the pseudo-spin) doesn't become excited in the transition.The second and the third peaks involve transitions corresponding to one and two pseudo-spin excitations, $\Omega_0$; these are recognized as interband transitions in the language of multiple band models. This first order perturbation approximation result remains valid as long as $t/\Omega_0 \ll1$ Comparisons with exact results will be shown below.

\subsection{Two electron sector}

The same procedure can be performed with the more difficult case of two electrons. 
In this case there are 16 basis states, and again they can be divided into 8 even and 8 odd states. We use a slightly different notation --- there are now states involving double occupation of a single site, and those involving only single occupation. The states with double occupation are
\bea
|d_1{\rangle_{e \atop o}}&=& \frac{1}{\sqrt{2}}[
c^{\dagger}_{a\up}c^{\dagger}_{a\dn}|2\rangle_a\otimes |0\rangle_b
\pm 
|0\rangle_a \otimes c^{\dagger}_{b\up}c^{\dagger}_{b\dn}|2\rangle_b] \nonumber \\
|d_2{\rangle_{e \atop o}}&=& \frac{1}{\sqrt{2}}[
c^{\dagger}_{a\up}c^{\dagger}_{a\dn}|2\rangle_a\otimes |\bar{0}\rangle_b
\pm 
|\bar{0}\rangle_a \otimes c^{\dagger}_{b\up}c^{\dagger}_{b\dn}|2\rangle_b] \nonumber \\
|d_3{\rangle_{e \atop o}}&=& \frac{1}{\sqrt{2}}[
c^{\dagger}_{a\up}c^{\dagger}_{a\dn}|\bar{2}\rangle_a\otimes |0\rangle_b
\pm 
|0\rangle_a \otimes c^{\dagger}_{b\up}c^{\dagger}_{b\dn}|\bar{2}\rangle_b] \nonumber \\
|d_4{\rangle_{e \atop o}}&=& \frac{1}{\sqrt{2}}[
c^{\dagger}_{a\up}c^{\dagger}_{a\dn}|\bar{2}\rangle_a\otimes |\bar{0}\rangle_b
\pm 
|\bar{0}\rangle_a \otimes c^{\dagger}_{b\up}c^{\dagger}_{b\dn}|\bar{2}\rangle_b]
\label{double}
\eea
where the `d' in the ket stands for `double', and the subscripts `1,2,3,4' simply enumerate the states, starting with the lowest, $|d_1\rangle$, where both pseudo-spins are in their ground states, The three other basis
states correspond to the first site have an excited pseudo-spin state, the second site having an excited pseudo-spin state, and both sites having excited pseudo-spin states, respectively.

Similarly, for the basis states involving only singly occupied sites, the `s' in the ket stands for `single'; these are
\bea
|s_1\rangle_{e \atop o}= \frac{1}{\sqrt{2}}[
c^{\dagger}_{a\up}|1\rangle_a\otimes c^{\dagger}_{b\dn} |1\rangle_b
\pm 
c^{\dagger}_{a\dn} |1\rangle_a \otimes c^{\dagger}_{b\up} |1\rangle_b] \nonumber \\
|s_2\rangle_{e \atop o}= \frac{1}{\sqrt{2}}[
c^{\dagger}_{a\up}|1\rangle_a\otimes c^{\dagger}_{b\dn} |\bar{1}\rangle_b
\pm 
c^{\dagger}_{a\dn} |\bar{1}\rangle_a \otimes c^{\dagger}_{b\up} |1\rangle_b] \nonumber \\
|s_3\rangle_{e \atop o}= \frac{1}{\sqrt{2}}[
c^{\dagger}_{a\up}|\bar{1}\rangle_a\otimes c^{\dagger}_{b\dn} |1\rangle_b
\pm 
c^{\dagger}_{a\dn} |1\rangle_a \otimes c^{\dagger}_{b\up} |\bar{1}\rangle_b] \nonumber \\
|s_4\rangle_{e \atop o}= \frac{1}{\sqrt{2}}[
c^{\dagger}_{a\up}|\bar{1}\rangle_a\otimes c^{\dagger}_{b\dn} |\bar{1}\rangle_b
\pm 
c^{\dagger}_{a\dn} |\bar{1}\rangle_a \otimes c^{\dagger}_{b\up} |\bar{1}\rangle_b].
\label{single}
\eea
The first has both pseudo-spin states in the ground state, with the three others having excited pseudo-spin states as in the case of the doubly occupied states. Note that an equally valid set of states combines $|s_2\rangle$ and 
$|s_3\rangle$ symmetrically and anti-symmetrically (as we did in the middle two basis states of Eq.~(\ref{3states}) for the three electron sector).

We want the unperturbed ground state to reside in the space of states involving only ground state pseudo-spin states.
Confining ourselves to this sector only, the ground state wave function is
\be
|\psi^{(2)}_0\rangle = a_0|d_1\rangle_e + b_0|s_1\rangle_o
\label{state2}
\ee
with:
\bea
{a_0}^2 = \frac{1}{2}\biggl(1-\frac{U/2}{\sqrt{(U/2)^2+4t^2S^2}}\biggl)\\
{b_0}^2 = \frac{1}{2}\biggl(1+\frac{U/2}{\sqrt{(U/2)^2+4t^2S^2}}\biggl).
\eea 
This will be an accurate ground state wave function as long as the pseudo-spin excitation energy remains the largest energy scale in the problem, i.e. $U \ll \Omega_0$, along with the restriction already used, $t \ll \Omega_0$.
When $U \ll 2tS$ the ground state consists of nearly equal amplitudes of the two basis states; on the other hand, when $U\gg 2tS$, $a_0 \sim 0$ and $b_0 \sim1$, and the singly occupied basis state dominates the ground state, as expected. With these same assumptions the two electron ground state energy is given as
\be
E^{(2)}_0 \approx -\Omega_0 + U/2 - \sqrt{(U/2)^2 + (2tS)^2}.
\label{e0_two}
\ee
\begin{figure}[tp]
\centering
\includegraphics[height=5.in,width=2.7in]{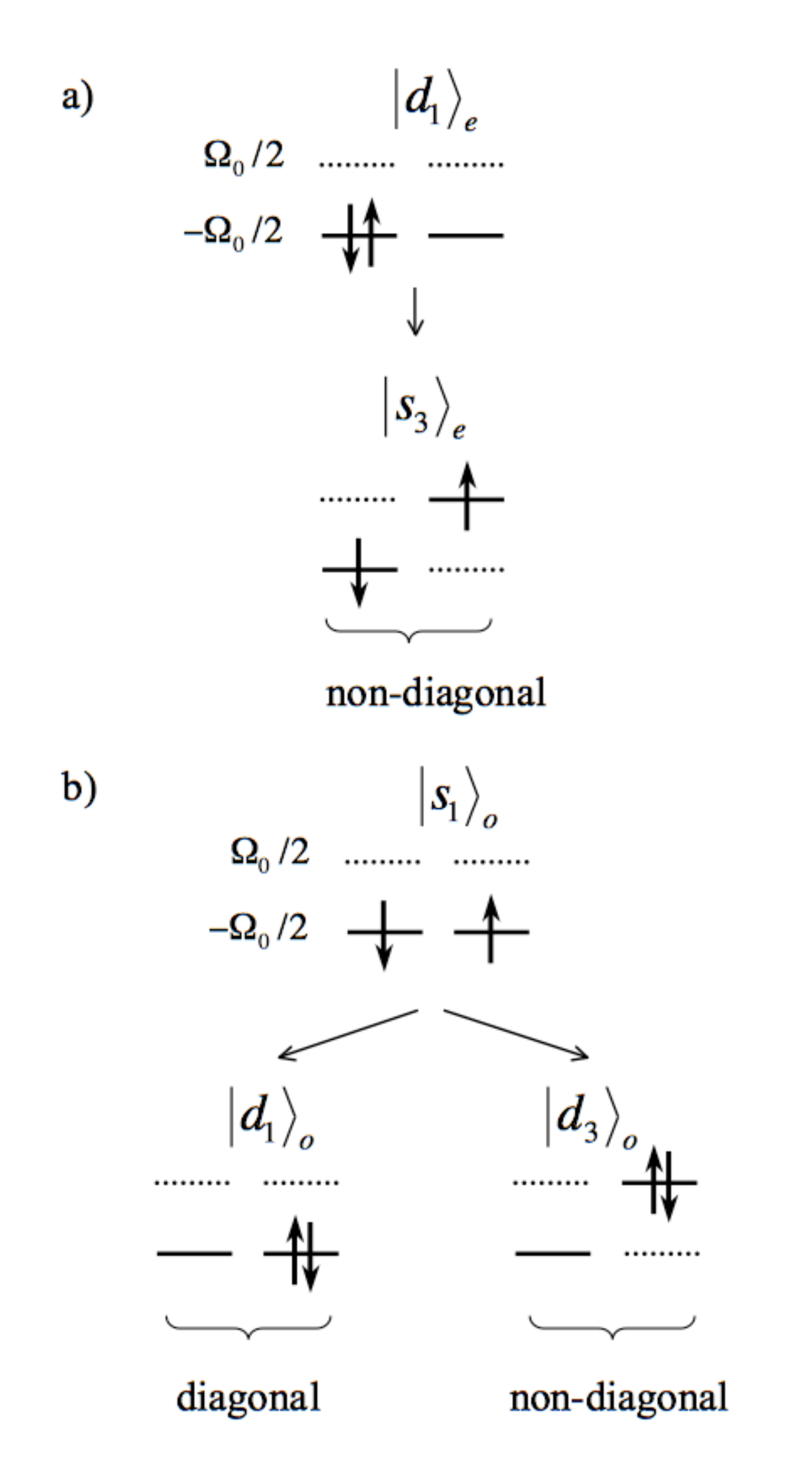}
\caption{Schematic depiction of the optical conductivity with two electrons in a dimer. Parts (a) and (b) refer to the
two basis states that make up the ground state wave function given by Eq.~(\ref{state2}) with ground state energy given by Eq.~(\ref{e0_two}). The first has only (two) non-diagonal transitions (the second transition, to states given by $|s_2\rangle_e$, is qualitatively the same as the one shown), while the second basis state has both diagonal and non-diagonal transitions, as shown. The state representations are schematic only; formulas in the text represent the full (even or odd) state.}
\label{2electron}
\end{figure}

To determine the optical conductivity we need the result of the current operator on each component of the ground state; the result is 
\bea
{J}|d_1\rangle_e &=& \frac{iet\bar{S}}{\hbar}\bigl(|s_2\rangle_e - |s_3\rangle_e \bigr) \nonumber \\
{J}|s_1\rangle_o&=& \frac{2iet}{\hbar}\bigl( S|d_1\rangle_o - \bar{S}|d_3\rangle_o \bigr).
\eea
Fig.~2 summarizes these transitions schematically. The unperturbed energies associated with these states are readily determined by inspection from Eqs.~(\ref{double}) and (\ref{single}). Using the analogue of Eq.~(\ref{conductivity}) appropriate to two electrons, with the ground state now a linear superposition of basis states given by Eq.~(\ref{state2}), the optical conductivity for two electrons is obtained as:
\bea
\sigma_1^{(2)}{(\omega)} = \frac{\pi e^2 t}{2 \hbar^2} &\Bigg\{& \frac{4tS^2}{\sqrt{(U/2)^2+(2tS)^2}}
\delta\bigl[\omega- \epsilon/\hbar \bigr] \nonumber \\
&+&\frac{4a_0^2 t \bar{S}^2}{\Omega_0 - U + \epsilon} \delta\bigl[\omega-\bigl(\Omega_0-U+\epsilon\bigr)/\hbar\bigr] \nonumber \\
&+&\frac{8b_0^2 t \bar{S}^2}{\Omega_0 + \epsilon} \delta\bigl[\omega-\bigl(\Omega_0+\epsilon\bigr)/\hbar\bigr],
\label{conductivity2}
\eea
where 
\be
\epsilon \equiv U/2 + \sqrt{(U/2)^2+(2tS)^2}
\label{epsilon}
\ee
plays the role of a low energy scale, i.e. $t$ (as $U \rightarrow 0$) or $U$ (as $t \rightarrow 0$). 
\begin{figure}[tp]
\centering
\includegraphics[height=3in,width=3.5in]{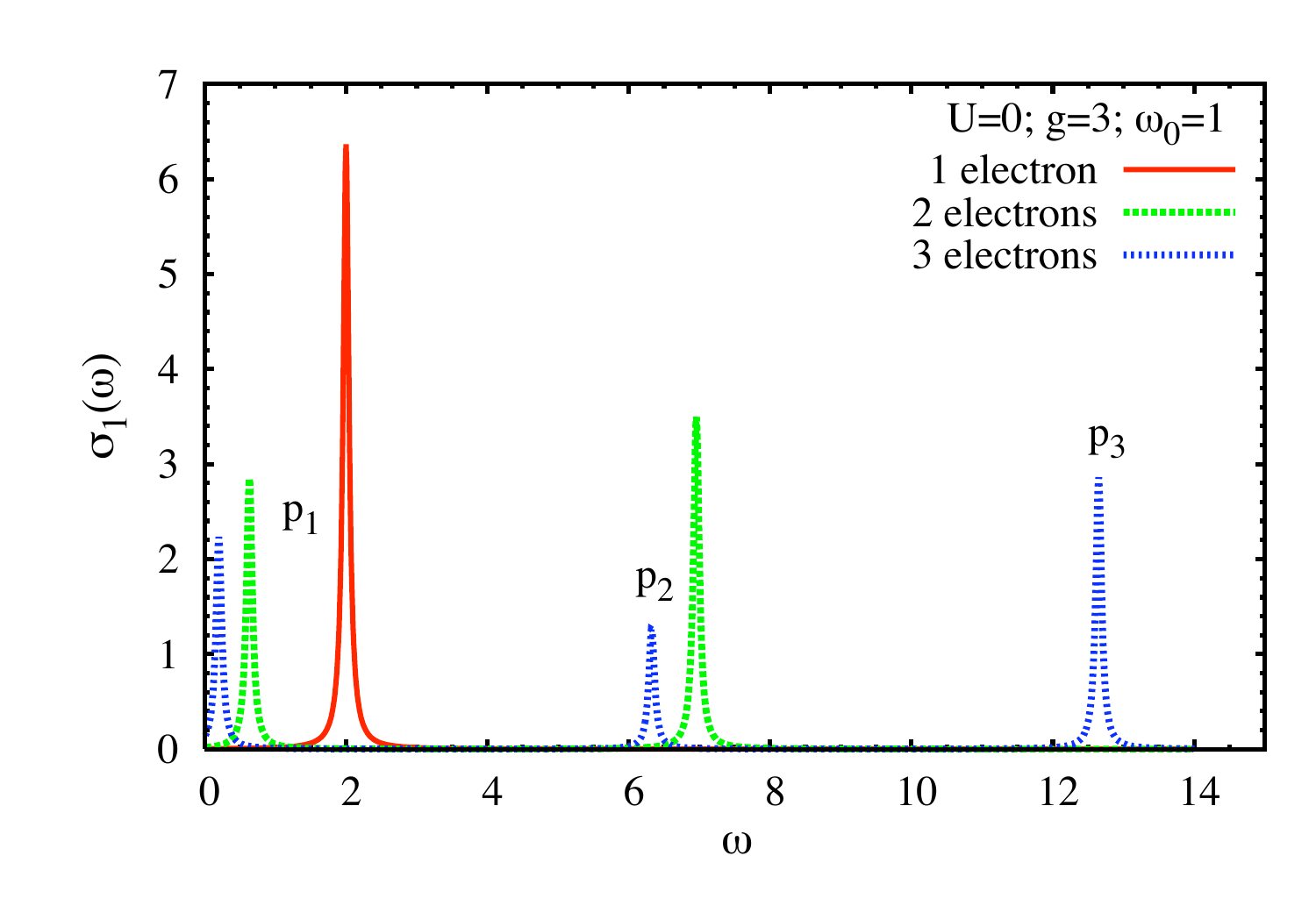}
\caption{The normalized optical conductivity as a function of frequency in a dimer obtained by perturbation theory with $U=0,g=3$ and $\omega_0=1$. The $\delta-$functions in the formulas in the text are represented here as broadened Lorentzians (with a width $0.05$). Note the decrease in relative low frequency spectral weight (labelled as $p_1$) as one goes from one to two to three electrons, indicating  a reduced mobility as the number of electrons increases. Also note as the number of electrons decreases the higher frequency relative weight decreases: for the two electron case spectral weight is entirely absent at $2\Omega_0$ (labelled as $p_3$) while for the one electron case weight is absent even for $\omega = \Omega_0$ (labelled as $p_2$).}
\label{opt5}
\end{figure}

\begin{figure*}[tp]
\centering
\includegraphics[height=7in,width=3.2in]{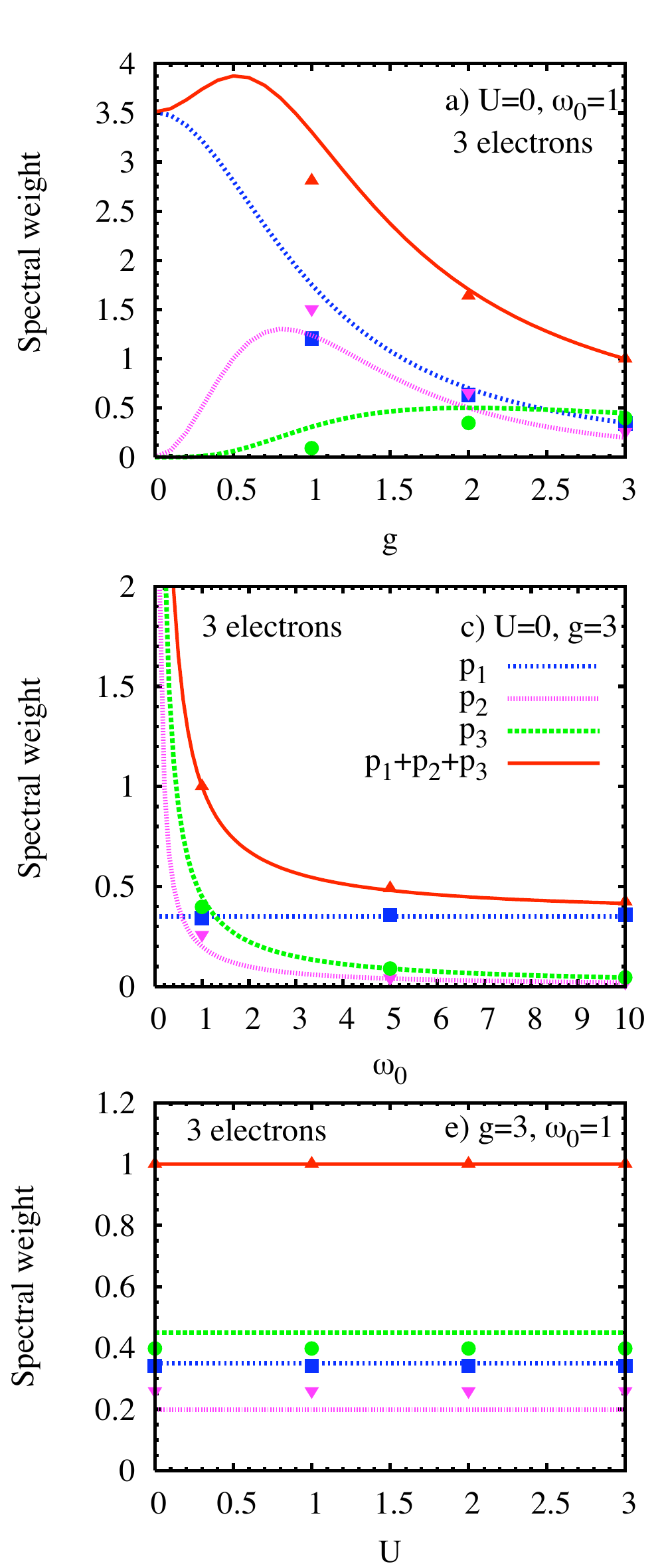}
\includegraphics[height=7in,width=3.2in]{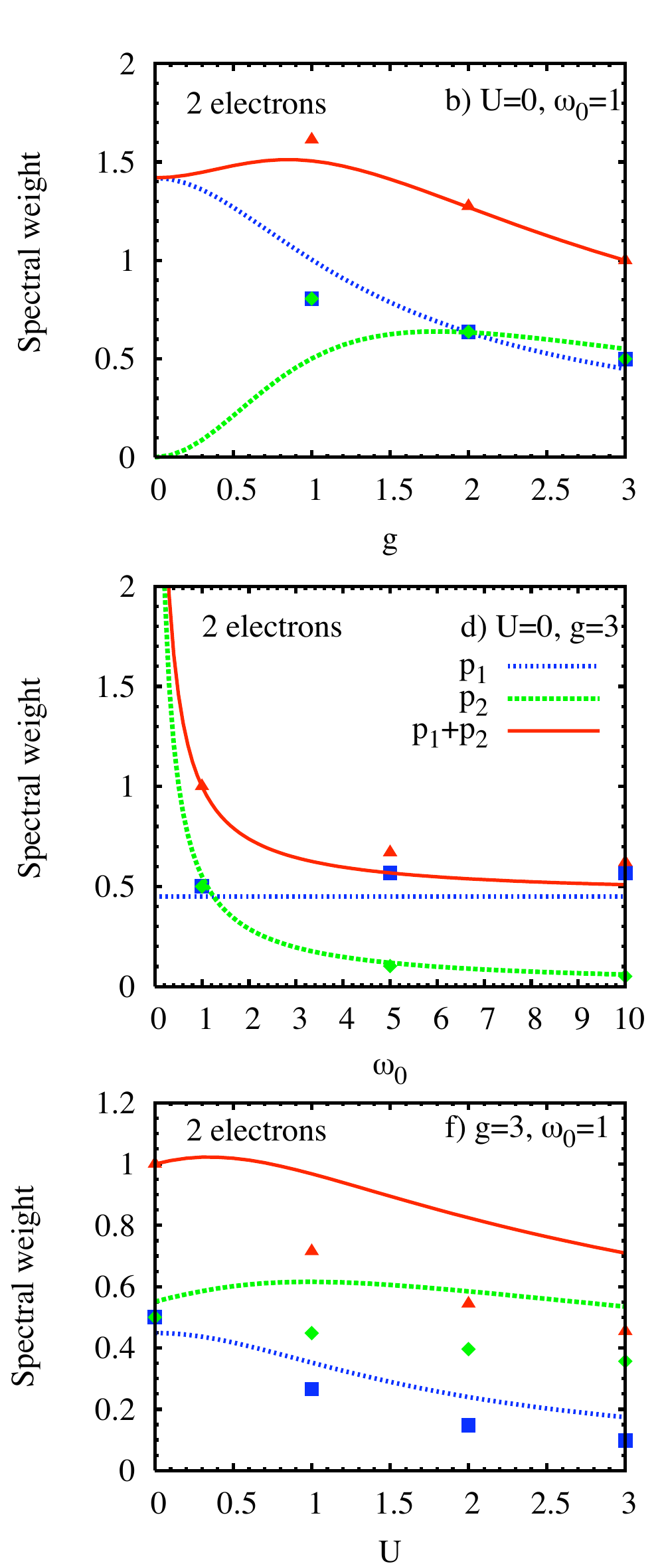}
\caption{The dependence of the various normalized spectral weight contributions to the optical conductivity on electron-pseudo-spin coupling $g$ for three electrons (a) and two electrons (b). In (c) and (d) we show the same quantities as a function of $\omega_0$, while in (e) and (f) they are shown as a function of $U$. Perturbation results are shown as curves, while exact results (for the dimer) are shown by symbols, as indicated. Note that usually the Drude weight dominates; however, for sufficiently large $g$ (and $U$ in the case of two electrons) the Drude weight is significantly reduced, indicative of reduced mobility, especially for the highly (electron) doped regime.}
\label{spt5}
\end{figure*}

The two electron optical conductivity also has three peaks, as seen in Eq.~(\ref{conductivity2}) or schematically in Fig.~2.
In the perturbative approach we have used, however, the two higher frequency peaks should be viewed as one occurring at a frequency scale of order $\Omega_0$ that has been split by a low energy scale of order $U$ or $t$. The analog of the third peak in the 3 electron case is absent here; there is no peak at $2\Omega_0$.

The first (low energy) peak corresponds to the diagonal transition from state $|s_1\rangle_o$ to $|d_1\rangle_o$, where, as in the three electron case, the hopping of an electron between sites occurs without modification of the pseudo-spin background. This diagonal transition corresponds to the Drude-like (or coherent) part of the optical conductivity, though it may extend to a range of (low) frequencies on the scale of $U$ as well as $t$ (as represented by $\epsilon$). 

The second peak at frequency $\omega = \Omega_0 - U + \epsilon$ is given by the two transitions: $|d_1\rangle_e$ to both $|s_2\rangle_e$ and $|s_3\rangle_e$. The third peak at frequency $\Omega_0 + \epsilon$ is obtained by the $|s_1\rangle_o$ to 
$|d_3\rangle_o$ state transition, as detailed in Fig.~2. The second and third peaks correspond to non-diagonal transitions, which means that the pseudo-spin background is excited in the transition; this in turn corresponds to transitions involving higher energy bands, not included in our starting Hamiltonian. There is also an overall factor of two enhancement because there are now two carriers instead of the one carrier present in the three electron case presented above (or the one electron case shown below).

One can again verify the conductivity sum rule; for this we need the ground state wave function given to first order in 
$t/\Omega_0$. Straightforward calculation\cite{remark1} gives
\bea
|\psi_0^{(2)}\rangle && \approx a_0|d_1\rangle_e + b_0 |s_1\rangle_0 + \nonumber \\
&&\frac{a_0 t \bar{S}}{\Omega_0 - U + \epsilon} \bigl( |s_2\rangle_0 + |s_3\rangle_0 \bigr) - \frac{2t\bar{S} b_0}{\Omega_0 + \epsilon} | d_3\rangle_e,\,\,
\label{two_el_wave_fn}
\eea
and an evaluation of the kinetic energy expectation value gives
\bea
-&&\langle\psi^{(2)}_{0}|{K}|\psi^{(2)}_{0}\rangle = \nonumber \\
&&\frac{4t^2S^2}{\sqrt{(U/2)^2 + (2tS)^2}} + \frac{4t^2\bar{S}^2a_0^2}{\Omega_0 - U + \epsilon} + \frac{8t^2\bar{S}^2b_0^2}{\Omega_0 + \epsilon},\,\,
\label{sum2}
\eea
again in agreement with the weights in Eq.~(\ref{conductivity2}). 

\subsection{One electron sector}

Finally, similar calculations in the one electron sector are particularly simple, because the single site pseudo-spin ground states for zero and one electron are identical. This means that there are no pseudo-spin excitations arising from application of the current operator to the single electron ground state. The optical conductivity for one electron is given by the simple expression,
\be
\sigma_{1}^{(1)}(\omega) = \frac{\pi e^2 t}{2\hbar^2} \delta(\omega-2t/\hbar),
\label{opt_oneelectron}
\ee
which contains only a Drude contribution, with no normalization (as required by the sum rule).

\section{Discussion}

We wish to show the relative contributions to the conductivity for the three different electron densities that we can access. Because this is a tight binding model it will not conserve total oscillator strength. This makes comparisons for different numbers of electrons and/or different parameter values difficult.\cite{hirsch92} Here, for a given number of electrons, we will normalize the conductivity to the overall spectral weight in the conductivity for that number of electrons.
Fig.~\ref{opt5} shows the two-site optical conductivity with one, two and three electrons, using $U=0$ and $g=3$, with $\omega_0 = 1$, as a `standard' set of parameters. While not necessarily realistic, they are chosen specifically to remove Mott complications at half filling; here the presence of $U$ will result in a significant {\em decrease} in low frequency spectral weight, in spite of the increase in number of available carriers.  These parameters will serve to illustrate the spectral weight transfer physics inherent in this model. At the same time, it is clear that the perturbation calculation is valid if  $\omega_0$ is large enough so that  $\Omega_0$ is much larger than $U,t$ as assumed above. 

{\bff The simplest case is clearly that of one electron. Eq.~(\ref{opt_oneelectron}) and Fig.~\ref{opt5} show that a single peak is present; it is located at $\omega = 2t/\hbar$, a non-zero
value only because we are dealing with a dimer, and not the thermodynamic limit. In the thermodynamic limit this would be a $\delta$-function at zero frequency, representing the Drude contribution. Normally this peak would be broadened through, for example, impurity scattering, but here (and even in the DMFT study of Ref.~[\onlinecite{bach10}]) it remains a $\delta$-function, broadened in Fig.~3 artificially by hand, so as to be visible. For small but non-zero electron densities $U$ would contribute as well, but for the most part this picture would remain unchanged. In particular, excited pseudo-spin states are essentially absent (see Fig.~3 in Ref.~[\onlinecite{bach10}], which shows that that the expectation value of the pseudo-spin operator is essentially its ground state value for low densities). Because the electron density is relatively dilute, there are very few optical transitions involving doubly occupied sites, and therefore it is not possible to excite the pseudo-spin excited state. For the dimer the only representative sector for this physics is the one electron sector (two electrons already constitutes a somewhat crowded lattice).} 

{\bff In contrast the three electron sector represents the most `crowded' situation for a dimer, while the two electron sector is somewhat in between, and, as mentioned previously, the absence of a Coulomb repulsion ($U=0$) aids to highlight the pseudo-spin physics, and suppress the Mott-related physics (which, from our point of view, is not essential, and will complicate the analysis). Referring to Fig.~3,}
note that the 3 electron conductivity has a significant relative contribution at high frequency ($2\Omega_0$); this is
entirely absent in the two electron conductivity --- it has been pushed down to lower frequency ($\Omega_0$). {\bff The reason for this is as follows: with three electrons, the ground state consists of a doubly-occupied and a singly-occupied site, with the respective pseudo-spins at each site in its ground state --- see the first of Eqs.~(\ref{3states}). An optical transition can result in one of the three states shown in Fig.~1; one of these, state $|4\rangle_e$, has two excited pseudo-spin states, corresponding to an energy $2\Omega_0$. One of these excitations comes from the site with a single electron --- before the transition this site was doubly occupied, and the ground state for this configuration required a pseudo-spin ground state {\em corresponding to two electrons}. Since one has left, there is now a component of the pseudo-spin which corresponds to an excited state {\em for the one electron configuration}. Similar remarks apply for the site that was previously singly occupied and is now doubly occupied. For two electrons this cannot happen --- see Fig.~2 and note the absence of an alternative involving two excited pseudo-spin state. This is because the pseudo-spin ground state is the same for an empty and singly-occupied site --- see Eq.~(\ref{comp}) or the first of Eqs.~(\ref{overlap}), where $T=1$.

This accounts for the peak structure for the various electron sectors in Fig.~3. The dimer calculations have an `all or nothing' character to them --- no high frequency ($2\Omega_0$) peak for the two electron sector, and not even an intermediate ($\Omega_0$) peak for the one electron sector. Of course in the DMFT calculations the changes from one electron density to another are smoothed out, as one can see in Fig.~13 of Ref.~[\onlinecite{bach10}]. The other feature that is apparent in Fig.~3 is the decrease of spectral weight in the relative Drude (low frequency) portion as one goes from the one electron to the two and then three electron sector. This is due to the polaron-like hopping renormalization already discussed. The relative weight of the Drude portion is indicative of the coherence of the carriers, so again, in the dilute limit, electrons can hop while the background pseudo-spin degree of freedom remains in the same ground state at both the site from which the electron hops, and at the site to which the electron hops, because only empty or singly occupied sites are involved. In the more crowded lattice limit (here represented by the three electron sector), doubly occupied sites are necessarily involved, and then the background pseudo-spin has to adjust according to whether a singly or doubly occupied site is involved.}

The progression of spectral weight with electron number is perhaps best exemplified by examining the conductivity formulas, Eqs.~(\ref{opt1},\ref{conductivity2},\ref{opt_oneelectron}) for three, two, and one electron(s), respectively, for $U=0$. Then the low frequency spectral weights are (omitting the common factor $\pi e^2 t/(2\hbar^2)$) $S^2$ for three electrons (but one hole carrier), $2S/2 = S$ for two electrons, and unity for one electron; these weights steadily increase by reducing the number of electrons, since $S  < 1$ always, and this illustrates the principle that holes are less mobile than electrons.

We analyze in more detail our results for the frequency dependence of the optical conductivity. The three electron optical conductivity has three distinct peaks from low to high frequency, one at $\omega \approx t$, one at $\omega \approx \Omega_0$, and one at $\omega \approx 2\Omega_0$, whose weights we denote $\mathrm{p}_1$, $\mathrm{p}_2$ and $\mathrm{p}_3$, respectively. In the two electron conductivity, there are again three peaks, but as explained above, the two high frequency ones are at the same characteristic frequency (identical if $U=0$), so we will combine the weight from these two and denote it as $\mathrm{p}_2$; we will continue to use $\mathrm{p}_1$ for the lowest frequency peak, {\bff and of course for the one electron conductivity, there is only a low frequency Drude-like peak, which we will also denote as $\mathrm{p}_1$}. In Fig.~\ref{spt5} we plot these weights to show how the spectral distribution of the optical conductivity varies with the strength of coupling $g$ for 3 (a) and for 2 (b) electrons; in (c) and (d) we show the corresponding results as a function of $\omega_0$, and in (e) and (f) results are shown for a variation of $U$. In all cases, the optical conductivity has been normalized to the total spectral weight for the parameters used in Fig.~\ref{opt5}, separately for each electron number. 

As expected, increasing the coupling strength $g$ between the electron and the background (pseudo-spins) reduces the mobility of the electron as obtained in the spectral weight of the first peak $\mathrm{p}_1$ (which would correspond to the Drude weight for an infinite lattice). {\bff This is simply due to the polaron effect mentioned above; with increased coupling, the amount of `background adjustment' required as the electron hops increases. Physically, the actual coupling in a given lattice is given by the amount of multi-orbital involvement required to minimize the energy locally when two electrons try to accommodate one another on the same site. Since we model this process with the pseudo-spin degree of freedom, we span a considerable parameter range in the figures.} The absolute weight $\mathrm{p}_2$ of the second peak for the three electron case is given analytically as $4t{g^2 \over (1+g^2)^2)}{t \over \Omega_0}$ (see Eq.~(\ref{opt1}), and achieves its maximum value at $g=\sqrt{2/3} \approx 0.8$, which is independent of $U$. For the two electron case, the off-diagonal transition contributions, represented by $\mathrm{p}_2$, are quite negligible compared with the low frequency weight $\mathrm{p}_1$ at weak coupling, but they play a more important role at strong coupling. Note that results arising from a complete diagonalization of all the dimer states are also shown, and, for these parameter regimes, the agreement is excellent, as expected. {\bff The transitions denoted by $\mathrm{p}_2$ and $\mathrm{p}_3$ represent incoherent processes; they may well correspond to the mid-infrared band that seems to feature so prominently in a wide variety of cuprate 
superconductors.\cite{tanner92} The other experimental feature to which we can make contact with these dimer calculations is the dependency on doping. Experimentally, the anomalies at the superconducting transition are most pronounced in the low hole regime,\cite{carbone06} consistent with the fact that the pseudo-spin physics in these calculations plays a large role precisely in this regime as well.}

{\bff Comparison with the results obtained from DMFT calculations\cite{bach10} is also possible. For example, in Fig.~13 of Ref.~[\onlinecite{bach10}], we show the conductivity as a function of frequency for various electron densities. Note that the parameters used in the DMFT calculation are in the more weak to intermediate coupling regime. Nonetheless, the calculations here are semi-quantitatively consistent with those. The first panel there refers to the very dilute limit ($n=0.1$), and as suggested here, there is a single low frequency Drude peak. Of course there it is centred around zero frequency, while here it is at $2t$, for reasons already explained. In the last panel in the same Fig.~13, $n=1.9$, corresponding more to our present three electron calculation. We expect
weight at $\Omega_0 = 2\omega_0\sqrt{1 + g^2} \approx 5.7$, which is very close to the one shown there. Furthermore the Drude-like peak has reduced spectral weight (clearer in Fig.~15 of Ref.~[\onlinecite{bach10}]) compared to the result at $n=0.1$. The expected peak at $2\Omega_0$ is, however, barely present, and at a higher frequency than expected. It is not clear what the cause of this latter discrepancy is, especially in light of the quantitative accuracy of the other peaks.}

For completeness we have included plots to show the variation with $\omega_0$ and $U$, where the expected behaviour occurs. Note that at half filling (two electrons) the exact results differ considerably from the perturbation theory results, as Mott physics becomes more prevalent (this is not surprising since this was not considered in the perturbative approach we took). 
{\bff As $\omega_0$ increases the results for three and two electrons become dominated by the Drude-like peak near the origin. Again, this is entirely expected, since pseudo-spin excitations become more and more energetically costly, and so, as seen explicitly in our perturbative expressions, energy denominators increasingly suppress these transitions requiring excited pseudo-spin states, so that these play much less of a role as $\omega_0$ increases. As is clear from panels (c) and (d), exact diagonalization results support these perturbative calculations.}


\section{Conclusions}

We have investigated spectral properties of the Dynamic Hubbard Model on a dimer, primarily to gain a qualitative understanding of the physics of electron-hole asymmetry, and polaron-like mobility inherent in real atoms. Primarily we have investigated the spectral features of the optical conductivity with different numbers of electrons. {\bff The physics we are trying to capture is that when electron movement results in a change from a doubly occupied site to a singly occupied site, or vice-versa, a considerable amount of `background' adjustment needs to take place. In real atoms this is apparent in that the orbitals occupied by a single electron are considerably modified when two electrons occupy that same orbital. In the Dynamic Hubbard model, these modifications are simulated by a pseudo-spin degree of freedom, at each site; an excited pseudo-spin state corresponds to an electron (partially) occupying an orbital that does {\em not} minimize the electron-ion energy, but does minimize the (local) electron-electron repulsion.

Such processes will impact the optical sum rule; in particular, weight will be transferred over a considerable range of energies, as a function of temperature and as a result of a phase transition. A considerable variation is expected as a function of electron concentration, and it is this aspect on which we have focused in the dimer calculations presented here.
If the electron concentration is low, the pseudo-spin degree of freedom will be rarely excited, and the electrons will be highly coherent. However, if the electron concentration is high,
then electron movement will be accompanied by pseudo-spin excitations. There is considerable experimental evidence for such incoherent processes in the cuprates, namely the mid-infrared band.
Our calculations clearly indicate that the Drude-like portion for holes has reduced mobility compared to that of electrons. The connection of the optical sum rule to the kinetic energy, and how this probe can demonstrate this physics has been worked out in great detail for the dimer system considered here. More detailed comparison to experiment will have to rely on DMFT calculations\cite{bach10} that provide answers in the thermodynamic limit.}

The results of the dimer calculations presented here agree with the physics originally obtained in a model in which the pseudo-spin degree of freedom impacted the on-site energy of an electron.\cite{hirsch92} Here, the pseudo-spin degree of freedom alters the effective Coulomb interaction between two electrons through a dynamical change in the on-site electron-electron interaction epitomized by $U$.\cite{hirsch01} The qualitative picture obtained here also provides a better
understanding of the conclusions obtained for an infinite lattice in Ref.~[\onlinecite{bach10}]: holes are less mobile than electrons, and the optical spectral weight distribution is significantly different for holes than for electrons.

\begin{acknowledgments}

We thank Jorge Hirsch for helpful discussions.
This work was supported in part by the Natural Sciences and Engineering
Research Council of Canada (NSERC), by ICORE (Alberta), and by the Canadian
Institute for Advanced Research (CIfAR).

\end{acknowledgments}

\bibliographystyle{prb}

\begin{thebibliography}{1}
\bibitem{remark1} Note that we have already diagonalized the Hamiltonian inside the two state subspace in which the pseudo-spin is always in the ground state. The excited state within this subspace is therefore omitted in the perturbation expansion.
\end{thebibliography}

\end{document}